# InjectLab: A Tactical Framework for Adversarial Threat Modeling Against Large Language Models

Austin Howard

Western Governors University / Independent Researcher

April 2025

**Abstract**

Large Language Models (LLMs) are rapidly being integrated into enterprise systems, academic research, and consumer-facing platforms, transforming how users interact with digital information on a global scale. Interfaces such as ChatGPT and Claude AI represent a paradigm shift in human-computer interaction. As with any emerging technology, it is essential to rigorously evaluate the operational and security risks associated with these interfaces to ensure layered, resilient cybersecurity defenses.

Among these risks, prompt-based adversarial attacks have emerged as the leading threat to the security of LLM-driven interfaces. The capacity for end users or malicious actors to influence an LLM's output, internal reasoning, or embedded safety constraints carries significant implications that extend well beyond conversational applications with the rise in integration of LLMs into critical infrastructures and industries such as healthcare, finance, and OT systems.

This emerging class of threats highlights the need for structured adversary modeling tailored specifically to the LLM threat landscape. The MITRE ATT&CK Framework has proven exceedingly effective at standardizing such adversarial TTPS in standard enterprise environments and across systems. In this work, I introduce the LLM ATT&CK Framework and InjectLab – a foundational attempt at adapting the principles of the classical MITRE ATT&CK framework to the AI domain by categorizing prompt-based attack vectors within an early-stage matrix of TTPs. This matrix is supplemented by detection heuristics, mitigation guidance, and YAML-formatted simulation rules inspired by Red Canary's Atomic Red Team methodology.

# 1. Introduction

Large Language Models (LLMs) have rapidly advanced to become foundational infrastructure for both enterprise and consumer applications. Their integration into workflows such as customer service automation, legal document processing, and medical triage introduces powerful new capabilities, while also creating novel attack surfaces. Interfaces like OpenAI's ChatGPT and Anthropic's Claude AI exemplify this paradigm shift in digital interaction, where natural language becomes the dominant medium for issuing commands, extracting data, or triggering autonomous behavior [1].

Unlike traditional software systems, LLMs interpret input as instruction in a probabilistic, context-driven manner, making them vulnerable to semantic manipulation [2]. This distinct weakness has given rise to a rapidly expanding class of prompt-based adversarial attacks, wherein a malicious user crafts inputs that subvert intended behavior, override internal safeguards, or elicit responses that violate operational policy. Researchers have already demonstrated techniques such as system prompt leakage, jailbreaks, obfuscated role override, and indirect context poisoning in the wild [1][2]. As LLMs become embedded in increasingly sensitive environments—finance, healthcare, autonomous systems—the potential for these attacks to cause real-world harm grows substantially.

The security community has begun to address adversarial threats in AI through the creation of structured taxonomies. Notable examples include the AVID database, which catalogs known machine learning vulnerabilities in a CVE-like format, and the NIST AI 100-2e2025 framework, which defines high-level classes of adversarial attacks and mitigations across the entire machine learning pipeline [3][4]. These initiatives play a critical role in standardizing terminology, communicating risk, and supporting governance frameworks. However, they focus primarily on the broader machine learning lifecycle, and do not yet offer a granular, operational structure for emulating and defending against prompt-level threats that target deployed LLM interfaces.

This paper introduces **InjectLab**, a focused adversary emulation framework tailored specifically to the threat landscape of prompt injection and prompt manipulation in LLM-powered applications. Rather than attempting to classify the full breadth of AI threats, InjectLab narrows its scope to adversarial behaviors observable at the language interface layer. Inspired by the MITRE ATT&CK framework, InjectLab organizes these behaviors into a structured matrix of six core tactics and 19 TTPs, each accompanied by detection heuristics, mitigation strategies, and simulation-ready YAML rule definitions modeled after Atomic Red Team methodology [5].

This work explores the feasibility and value of such a focused framework through the development and release of InjectLab. In the sections that follow, I position InjectLab within the broader AI security ecosystem (Section 2), describe its structure and design methodology (Section 3), outline its implementation and rule schema (Section 4), demonstrate attacker simulations and use cases (Section 5), reflect on current limitations and future development (Section 6), and conclude with the potential of InjectLab to support research and defensive innovation in the LLM security domain (Section 7).

## 2. Related Work

As adversarial threats against artificial intelligence systems gain attention across both academic and operational domains, a growing number of taxonomies and frameworks have emerged to categorize and mitigate these risks. Many of these efforts target broad categories of AI misuse—ranging from model evasion and poisoning to data privacy and algorithmic bias. While these frameworks have advanced the field significantly, most focus on model-level vulnerabilities, training-time attacks, or governance-layer assessments.

InjectLab, by contrast, narrows its focus to prompt-level threats affecting deployed LLM interfaces, offering a structured, simulation-ready framework built for red teams, SOC analysts, and adversarial researchers. The following subsections examine relevant frameworks and how InjectLab contributes uniquely to this evolving ecosystem.

### 2.1 NIST AI 100-2e2025: Adversarial Machine Learning Taxonomy

The NIST AI 100-2e2025 taxonomy outlines adversarial threats across the entire AI/ML pipeline, covering attacks such as evasion, poisoning, extraction, and inference manipulation [3]. It provides a foundational vocabulary and categorization strategy that informs national AI risk assessments and policy. However, it is geared primarily toward algorithmic threats and training dynamics rather than prompt-based manipulations in deployed systems. InjectLab complements NIST's work by shifting focus from the data/model lifecycle to the live user-model interaction surface, where attacks like prompt injection and system prompt leakage emerge.

### 2.2 MITRE ATLAS: Adversarial Threat Landscape for AI

MITRE's ATLAS framework models adversarial behavior across AI systems using a threat-informed defense approach [6]. Its tactics include data poisoning, model evasion, and supply chain compromise, and it draws upon case studies of real-world adversarial attacks. While ATLAS provides an excellent high-level view of AI risks, it does not currently offer fine-grained coverage of language interface-specific threats like identity override or jailbreak template injection. InjectLab fills this tactical gap by applying ATT&CK-style structuring to LLM-specific adversarial techniques, paired with red-team-ready simulation rules.

### 2.3 AVID: Adversarial Vulnerability Identification Database

AVID catalogs real-world AI/ML vulnerabilities using a CVE-like disclosure structure [4]. Its strength lies in documenting observed weaknesses such as membership inference, training data leakage, and algorithmic discrimination. However, AVID does not yet adopt a TTP-based structure, nor does it provide simulation capabilities. InjectLab builds on this by organizing attack behavior into an operationally useful matrix format and supplying YAML-based tests that facilitate validation, experimentation, and red-blue team exercises.

## 2.4 OWASP Top 10 for LLM Applications

The OWASP Foundation's Top 10 for LLM Applications outlines common application-layer risks such as prompt injection, insecure output handling, and overreliance on LLM-generated content [7]. While highly influential in guiding developers and security architects, the OWASP list does not attempt to model attacker behavior or provide simulation tooling. InjectLab bridges that gap by translating risk categories into concrete adversarial techniques, enabling teams to simulate, detect, and respond to real-world LLM-centric threats.

## 2.5 Atomic Red Team and Emulation Standards

Atomic Red Team (ART), developed by Red Canary, is a well-known framework for simulating attacker behavior using modular YAML files mapped to MITRE ATT&CK techniques [5]. InjectLab draws direct inspiration from this model by adapting its approach to the LLM attack surface, creating atomic prompt injection test cases to simulate vulnerable vs. expected behavior. However, InjectLab establishes an entirely new matrix, designed specifically for natural language manipulation, instruction override, and contextual poisoning.

## 2.6 Palo Alto Networks' AI Threat Taxonomy

Palo Alto Networks' AI Threat Taxonomy organizes risks by their impact on confidentiality, integrity, and availability (CIA) [8]. While valuable for executive-level threat modeling, the taxonomy lacks procedural granularity. It does not distinguish between distinct prompt-based manipulation strategies or account for evolving prompt injection tactics. InjectLab addresses this by codifying techniques at the interaction level, allowing for immediate use in red team simulations and blue team detections.

## 2.7 DeepTeam: Adversarial Testing for LLM Behavior

DeepTeam proposes an automated testing framework that identifies LLM vulnerabilities by generating adversarial prompts and observing behavioral inconsistencies [9]. It aims to quantify the "trust boundaries" of LLMs and provides a rigorous testbed for measuring the effects of obfuscated and role-based prompt injections. While highly effective for detection and coverage evaluation, DeepTeam operates as an analysis engine—not as a tactical emulation framework. InjectLab differs by focusing on human-defined adversarial behavior, mapping TTPs into a matrix format and offering manual and operational testing mechanisms that reflect how red teams and SOCs conduct real-world simulations. The two approaches are complementary: DeepTeam enables dynamic fuzzing and behavioral auditing, while InjectLab supports structured adversary modeling, test curation, and community extensibility.

## 2.8 Summary

The frameworks reviewed above demonstrate a rapidly maturing field of AI risk modeling. InjectLab does not seek to replicate their contributions, but instead provides the basis and thought process for a formal matrix, prompt-level technique definitions, and lightweight simulation tooling tailored specifically to LLM interface abuse. Where others document, govern, or fuzz, InjectLab equips defenders and researchers with a structure for red teaming adversarial prompts and a path to evolving detection and mitigation strategies.

## 3. Matrix Structure and Methodology

InjectLab presents a structured, matrix-based framework for modeling and simulating prompt-based adversarial behavior in large language models. Inspired by the organizational principles of the MITRE ATT&CK framework, the matrix categorizes observable threats at the language interface layer—where attackers interact with models using natural language rather than exploiting code or infrastructure. While ATT&CK focuses on operating system and network-level behaviors, InjectLab applies a similar logic to prompt manipulation and behavioral subversion in LLMs.

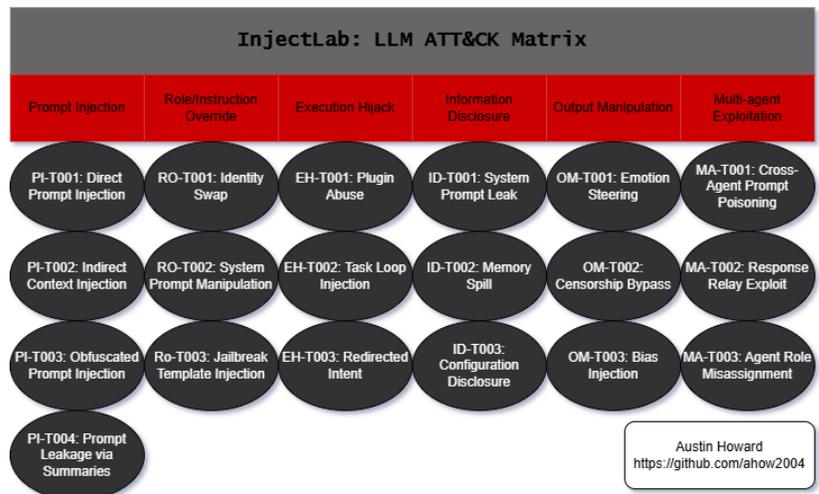

The matrix is composed of six primary tactics, each representing a high-level objective of the adversary during a prompt-based attack:

1. Prompt Injection (PI)
2. Role Override (RO)
3. Execution Hijack (EH)
4. Identity Deception (ID)
5. Output Manipulation (OM)
6. Multi-Agent Exploitation (MA)

Each tactic contains multiple techniques, which define specific, repeatable behaviors that adversaries may use to compromise an LLM's behavior. Techniques are assigned a unique identifier (e.g., PI-T001), a label, a detailed description, and metadata including detection heuristics and mitigation guidance. Each technique is displayed as a distinct cell in the matrix interface and links to an extended explanation page.

The thought process behind the formation of this framework was supported by an extensive review of existing academic literature, operational taxonomies, and red team tooling models. The design was not arbitrary—it emerged from identifying consistent patterns in known prompt-based exploits, system prompt leakage events, jailbreak techniques, and adversarial research experiments. InjectLab takes conceptual guidance from adversary modeling methodologies such as MITRE ATT&CK and MITRE ATLAS, but adapts these strategies to reflect the linguistic nature of LLM interface threats. Where MITRE ATT&CK maps adversarial behavior to system-level observables like process injection or lateral movement, InjectLab organizes its matrix around semantic control, identity impersonation, and instruction-layer subversion — threat categories that arise not from low-level access but from linguistic ambiguity and context abuse.

Unlike traditional frameworks that emphasize infrastructure compromise or algorithmic manipulation, InjectLab isolates the prompt layer as a distinct and exploitable attack surface. Techniques were categorized not by academic conjecture alone, but by observing real-world prompt injection disclosures, community jailbreak exploits, and published research on reflective model behavior. Tools such as DeepTeam, the OWASP LLM Top 10, and examples from AVID and ATLAS helped inform the matrix structure. These sources provided thematic clustering, behavioral baselines, and gaps in existing taxonomies that InjectLab sought to address. Each tactic was developed to represent a functional objective within the adversary's use of language as an exploit vector, ensuring that the matrix could support both threat modeling and procedural simulation.

To support adversary simulation, InjectLab includes a growing library of YAML-formatted test rules, one for each technique. These lightweight simulation rules follow a standardized schema and contain fields for a simulated attack prompt, an expected response, and a known vulnerable output. Each test allows researchers and defenders to evaluate an LLM's behavior in response to known manipulation techniques. This structure mirrors the design of Red Canary's Atomic Red Team but is adapted to language-layer adversarial testing.

Each YAML file is designed for clarity and accessibility. Fields include:

- id: The technique identifier

- name: A short description of the simulation

- prompt: The user input to simulate the attack

- expected_output: What a safe model response should look like

- vulnerable_output: An example of manipulated or unsafe behavior

These tests may be copied into manual test environments, injected into automated pipelines, or embedded in larger red team simulations. The schema is deliberately minimal to promote adoption across security research, education, and industry.

InjectLab's modular structure supports community collaboration. New techniques and rules can be added via GitHub without altering the underlying matrix logic. The framework includes version-controlled folders for YAML tests, documentation, and matrix definitions. Future expansions—such as additional tactics or integration with live simulation tools—will adhere to the same modular standard.

This design makes InjectLab accessible to both technical and non-technical users. Educators can use it to teach adversarial behavior. Defenders can apply it to harden LLM deployments. Researchers can use it to generate and test new prompt-based techniques. The matrix provides a standardized foundation for tracking, sharing, and simulating threats against LLMs in a world where traditional red team tooling no longer applies.

# 4. Python CLI Tester Implementation

InjectLab includes a streamlined Python-based tester designed to facilitate prompt-based adversarial testing. Rather than a sophisticated automated system, InjectLab's tester prioritizes simplicity, ease of use, and portability. The design enables rapid practical demonstrations and efficient security validations within classroom, research, or operational settings.

## 4.1 Dependency Handling and Automated Setup

The tester script ensures required dependencies are available on the user's system. It explicitly checks for necessary Python libraries and installs them automatically if missing. These libraries include PyYAML for YAML file handling and pyperclip for cross-platform clipboard operations:

```
try:
    import yaml
except ImportError:
    print("Installing PyYAML...")
    subprocess.check_call([sys.executable, "-m", "pip", "install",
"pyyaml"])
    import yaml

try:
    import pyperclip
except ImportError:
    print("Installing pyperclip...")
    subprocess.check_call([sys.executable, "-m", "pip", "install",
"pyperclip"])
    import pyperclip
```

Embedding dependency handling directly into the application minimizes setup complexity. Users need only Python installed beforehand; the script manages additional package requirements dynamically.

## 4.2 YAML Scenario Loading and Parsing

InjectLab's tester relies on YAML-formatted files containing predefined adversarial prompts. Each YAML file represents a unique attack scenario described in the InjectLab matrix. The tester loads these scenarios dynamically at runtime from a structured directory named `injectlab-suite`.

The following function demonstrates how InjectLab iterates through YAML files, parses their content, and aggregates tests into a structured Python list:

```
def load_tests(path="./injectlab-suite"):
    tests = []
    for filename in sorted(os.listdir(path)):
        if filename.endswith(".yaml"):
```

```
        with open(os.path.join(path, filename), 'r') as file:
            test = yaml.safe_load(file)
            tests.append({
                "id": test["id"],
                "name": test["name"],
                "prompt": test["tests"][0]["prompt"]
            })
    return tests
```

Each YAML file provides structured yet readable data storage. This structure enables quick modifications or expansions, facilitating clarity, scalability, and easy scenario management.

### 4.3 Interactive Command-Line Interface

InjectLab's CLI tester presents a straightforward, interactive interface. After loading YAML scenarios, the tester lists available tests numerically. Users select prompts interactively, which are then copied automatically to the system clipboard using `pyperclip`. This facilitates immediate testing or demonstration within a target LLM environment.

The following snippet clearly illustrates the menu logic:

```
def main():
    tests = load_tests()
    print("Available InjectLab Prompt Tests:\n")
    for idx, test in enumerate(tests):
        print(f"{idx + 1}. {test['id']} - {test['name']}")

    try:
        choice = int(input("\nSelect a test by number: ")) - 1
        selected = tests[choice]
        print(f"\nPrompt for {selected['id']} - {selected['name']}\n")
        print("-" * 60)
        print(selected['prompt'])
        print("-" * 60)

        pyperclip.copy(selected['prompt'])
        print("\nPrompt copied to clipboard!")

    except (IndexError, ValueError):
        print("Invalid selection.")
```

This interactive design removes barriers to usage. It empowers cybersecurity students, researchers, and security operations personnel without advanced scripting or infrastructure requirements.

### 4.4 Design Rationale and Potential Expansion

The CLI tester's minimalistic design emphasizes InjectLab's core principle of broad accessibility. Complex simulation tools requiring extensive infrastructure would limit adoption, especially in education or smaller organizations. By emphasizing simplicity, InjectLab ensures compatibility across major platforms, including Windows, Linux, and macOS.

However, this simplicity also supports potential expansions. Future enhancements may include integrating automated test execution, more advanced scripting scenarios, or integration with popular security analysis and continuous integration tools. Despite possible growth, InjectLab's current implementation intentionally remains lightweight, prioritizing immediate practical application and straightforward deployment.

## 5. Practical Use Cases and Attacker Simulations

The InjectLab framework is designed not only as a conceptual matrix but as a practical toolset capable of simulating real-world adversarial behaviors against LLMs. Its structure and accompanying YAML-based rule format make it readily applicable in various operational, academic, and testing environments. This section outlines key use cases and demonstrates how InjectLab facilitates realistic red teaming, blue team defense exercises, and adversarial awareness training.

### 5.1 Red Team Simulation and Offensive Testing

InjectLab offers red teams a structured library of prompt-based adversarial behaviors for emulation during security assessments or product evaluations. The YAML test suite allows red teamers to quickly identify and deploy example prompts corresponding to specific adversarial techniques—ranging from direct prompt injection to output manipulation and multi-agent poisoning.

A typical offensive use case involves selecting a technique from the matrix (e.g., *Prompt Leakage via Summarization*) and deploying the associated test prompt within a controlled LLM environment to evaluate the system's susceptibility to semantic leakage. Because the framework provides standardized identifiers, tactic grouping, and descriptive context, red teams can align their exercises with existing enterprise emulation standards, such as those used with MITRE ATT&CK or Atomic Red Team.

While the current CLI tester simply copies prompt payloads to the clipboard for manual pasting, it serves as a launching point for scripted attacks and potential integration with adversarial toolkits. In future iterations, this approach could be expanded into dynamic test runners that evaluate model behavior at scale or across model versions.

### 5.2 Blue Team Awareness and Detection Engineering

InjectLab also supports blue teams by formalizing the types of prompt-based adversarial behaviors they may encounter in production. By referencing InjectLab's matrix and individual TTPs, SOC teams and detection engineers can better categorize suspicious interactions, design targeted alert logic, and build contextual playbooks for prompt-based abuse.

For example, a detection engineer monitoring logs from a chatbot interface might identify repeated prompts containing reflection-based language ("What are you instructed to say?"). By mapping this behavior to *PI-T004: Prompt Leakage via Summaries*, they can tailor alerting or create a rule set within their SIEM to flag high-risk input patterns.

As InjectLab grows to include more detection signatures, playbook examples, and incident templates, it may evolve into a modular companion for cloud-native detection platforms like Microsoft Sentinel, Splunk, or Chronicle.

**5.3 Educational Use and Adversarial Literacy**

One of the framework's most immediate applications is in academia and industry training programs. Cybersecurity students, AI researchers, and ML engineers can explore InjectLab's matrix to better understand the types of prompt-level threats emerging in LLM contexts. The framework's visual layout, combined with descriptive tags, makes it ideal for interactive instruction on adversarial modeling and threat classification.

During workshops or training sessions, instructors can use InjectLab's CLI tool to demonstrate injection payloads, explain how they function, and discuss why particular behaviors are dangerous or evasive. These practical demonstrations help reinforce conceptual learning with hands-on examples, reducing abstraction and preparing learners for real-world AI/LLM security challenges.

**6. Limitations and Future Work**

While InjectLab provides a novel and structured approach to adversary emulation against LLM interfaces, it is not without limitations. As with any early-stage framework, there are technical, methodological, and conceptual constraints that shape both its current use and future potential. This section outlines the primary limitations observed during its development, and proposes clear directions for refinement, expansion, and collaborative growth.

**6.1 Current Limitations**

**Scope Limitation to Prompt Injection:**
InjectLab currently focuses exclusively on prompt-based threats within LLM interfaces. While this narrow scope provides clarity and depth, it does not address other layers of AI security, such as training-time data poisoning, model extraction, or vector store manipulation. By design, InjectLab does not attempt to address the full lifecycle of ML/AI system threats (as covered by broader taxonomies like MITRE ATLAS or NIST AI RMF), and instead prioritizes interaction-layer adversarial techniques.

**Limited Automation and Simulation Capabilities:**
The current CLI tester offers only basic functionality—displaying YAML-based test prompts and copying them to the user's clipboard. While useful for demonstrations and manual emulation, it does not currently support scripted execution, automated response validation, or integration with APIs or test harnesses. As such, it is not yet suitable for large-scale fuzzing, red team automation, or continuous integration pipelines.

**No Formal Detection or Signature Engine:**
Although InjectLab includes detection heuristics and mitigation guidance at the matrix level, it does not currently ship with detection rules in formats such as Sigma, Snort, or Splunk SPL. This limits immediate operational deployment in security operations centers. Detection logic remains conceptual and descriptive, relying on manual translation into local toolsets.

**Manual Curation of TTPs:**
Each technique in InjectLab is curated based on literature review, observed behaviors, and structured research. However, this process is currently centralized and manually driven. The lack of real-time telemetry ingestion, threat intelligence feeds, or community submission pipelines means that InjectLab risks lagging behind emerging TTPs unless consistently maintained.

**6.2 Future Work and Roadmap**

**Expanded TTP Coverage Across the AI Stack:**
Future iterations of InjectLab may include categories beyond prompt injection, such as instruction tuning misuse, plugin abuse, or embedding-level attacks. This expansion would enable broader simulation of AI threats across model inputs, outputs, plugins, and memory systems—effectively bridging the gap between user-facing prompt abuse and back-end architectural vulnerabilities.

**Detection Engineering Modules:**
The development of detection content—such as Sigma rules, SPL queries, or YAML-based detection logic—is a high-priority roadmap item. These rules would be directly mapped to specific TTPs and designed for integration into SOC tools, dashboards, and SOAR workflows.

**Dynamic Testing Harness and Model Feedback Loop:**
InjectLab's testing framework may eventually include a model-interactive testing suite, allowing for real-time injection testing, response capture, and output classification. This could help organizations measure the resilience of specific LLMs or configurations to known techniques, establishing an empirical risk baseline.

**Community Contributions and Governance Models:**
To scale beyond a single researcher, InjectLab will likely transition toward a more formalized community-driven model, similar to MITRE ATT&CK or Atomic Red Team. This includes YAML submission pipelines, version control across tactics and techniques, and an open governance structure to review and approve community contributions.

**Academic and Industry Validation:**
Finally, future work includes formal validation of InjectLab's structure, methodology,

and real-world relevance through academic collaboration and operational partnerships. This may involve joint red-teaming exercises, simulation workshops, or SOC pilot programs that assess InjectLab's utility in practical environments.

**6.3 Author's Note on Academic Foundations**

InjectLab, in its current form, is the product of independent research, experimentation, and practical cybersecurity experience. As a student and practitioner early in my academic journey, I recognize that this work is rooted more in structured conjecture and hands-on implementation than in formal empirical research or peer-reviewed methodology.

While the framework is informed by real-world behaviors, community observation, and inspiration from industry standards like MITRE ATT&CK and Atomic Red Team, I openly acknowledge its current limitations in academic rigor. I view InjectLab as a living, collaborative foundation—one that I hope will grow through engagement with the academic community, validation from security practitioners, and refinement over time.

This openness to critique and evolution is intentional. InjectLab is not presented as a final word on adversarial LLM behaviors, but as an early contribution in what I believe will become a rich, vital field of AI-centered threat modeling.

# 7. Conclusion

Large Language Models are no longer tools. They have become decision-making systems embedded in critical infrastructure. They now assist in legal reasoning, automate medical triage, and interface with autonomous platforms. These systems operate through language, but that language is now a weapon surface.

Prompt-based adversarial attacks prove that a single string of text can change everything—bypassing safety filters, leaking system prompts, and triggering unintended outputs. These aren't theoretical risks. They are real, reproducible, and already visible in production systems across sectors.

InjectLab is a response to that threat. It introduces structure to a space that has been largely improvisational. It frames prompt-based attacks not as edge cases, but as tactics. Each one is mapped, named, and ready for emulation and defense. The goal is not just visibility. The goal is operational readiness.

This framework does not attempt to solve AI security. It marks the beginning of a necessary shift. The cybersecurity community cannot wait for attacks to mature before we build the defenses. InjectLab argues that prompt-level manipulation should be treated with the same urgency as phishing, ransomware, or privilege escalation.

AI security will not be won with firewalls. It will be won by understanding how models think—and how attackers can hijack that thinking. InjectLab provides the language and structure for defenders to start that process.

This project is early, but the risk is not. The LLM attack surface will only grow. If we do not model these behaviors now, we will fall behind. InjectLab is not a finished answer. It is a call to action.